\documentclass[12pt]{iopart}

\usepackage{iopams}  
\usepackage[utf8]{inputenc}
\usepackage{setstack}
\usepackage{amsfonts}
\usepackage{amssymb}
\usepackage{graphicx}
\usepackage{float}
\usepackage{relsize}
\usepackage[backend=bibtex]{biblatex}
\usepackage{biblatex}
\usepackage[left=3cm,right=3cm]{geometry}

\begin{document}

\title[ ]{Time Evolution of Gaussian Wave Packets under Dirac Equation with Fluctuating Mass and Potential}

\author{Atis Yosprakob, Sujin Suwanna}

\address{MU-NECTEC Collaborative Research Unit on Quantum Information, Department of Physics, Faculty of Science, Mahidol University, Bangkok, Thailand, 10400.}

\ead{atis.yos@student.mahidol.ac.th, sujin.suw@mahidol.ac.th}

\vspace{10pt}
\begin{indented}
\item[]May 2016
\end{indented}

\begin{abstract}
Localization of relativistic particles have been of great research interests over many decades. We investigate the time evolution of the Gaussian wave packets governed by the one dimensional Dirac equation. For the free Dirac equation, we obtain the evolution profiles analytically in many approximation regimes, and numerical simulations consistent with other numerical schemes. Interesting behaviors such as Zitterbewegung and Klein paradox are exhibited. In particular, the dispersion rate as a function of mass is calculated, and it yields an interesting result that super-massive and massless particles both exhibit no dispersion in free space. For the Dirac equation with random potential or mass, we employ the Chebyshev polynomials expansion of the propagator operator to numerically investigate the probability profiles of the displacement distribution when the potential or mass is uniformly distributed. We observe that the widths of the Gaussian wave packets decrease approximately with the power law of order $o(s^{-\nu})$ with $\frac{1}{2}<\nu<1$ as the randomness strength $s$ increases. This suggests an onset of localization, but it is weaker than Anderson localization.
\end{abstract}

%
%
%
%
%

\section{Introduction}

We are interested in investigating a question concerning the propagation of a relativistic particle governed by the Dirac equation. This question has been of great interests over many decades as it related closely to phenomena such as the Klein paradox and the Zitterbewegung \cite{Bosanac,Dombey199941}, but it has also been revived with the discovery of electron behaviors in graphene in recent years \cite{Nature620,PhysicsWorld,Nature438197,Allain,PhysRevA.84.052102,Nature438201}. There the electrons move much faster than those in ordinary conductors, and behave like relativistic particles having virtually no rest mass. Because of this, the electrons in graphene are governed by the Dirac equation, which exhibits a ballistic motion in two dimensions, and can potentially be used to test the aforementioned phenomena of Klein paradox and Zitterbewegung \cite{Nature620,PhysicsWorld} without requiring high energy.
 
In a broader sense, the propagation nature of relativistic quantum particles raises interesting questions. It is well-known that classical electromagnetic and electronic waves can undergo transitions between ballistic, diffusion and localization regimes when there is fluctuation, turbulence or randomness in the environment; see Refs. \cite{PhysRevLett.95.020401,PingSheng, Soukoulis, Ziegler20031189} and references therein. Is it possible and under what conditions that a quantum relativistic particle exhibits a localized wave function? For electrons with low energy in traditional conductors, their matter-waves propagation is governed by the non relativistic Schr\"odinger equation, so that the existence of fluctuation, turbulence or potential disorder in general will halt their propagation, and make their wave functions localized, exhibiting a phenomenon known as Anderson localization \cite{PhysRev.109.1492,Filoche11092012,Melloy,PhysRevLett.100.094101}. Much is known about the Anderson localization, and its corresponding metal-insulator transition. For examples, in one dimension, the existence of randomness always make the electronic wave function localized; while in three dimensions or higher, localization have been shown for sufficiently large disorder strength and extreme energies; see Ref.\cite{Suwanna} and references therein. In two dimensions, it is expected that localization will occur in the presence of randomness, much like in one dimension, but perhaps with a weaker decay of the wave function. 

For high energy or relativistic particles, there have been fewer results available in literature. In this situation, the wave propagation is governed by the Dirac equation, and the randomness can come in the form of mass or potential fluctuation \cite{NatureComm2384,Lock,Ramola,PhysRevE.91.042109}, possibly due to a variety of sources of fluctuation, such as the geometrical frustration of the system in the case of electrons in graphene, or random forces in other situations. Likewise, Anderson localization of light has received considerable attention recently \cite{PhysRevB.78.235101,PhysRevLett.85.4269,PhysRevA.64.053804,PhysRevLett112,Nature390}. There have been some reports which indicate both the existence of Anderson localization of light, and the absence of Anderson localization of light \cite{PhysRevA.79.032112,PhysRevLett112}. To our knowledge, results on the propagation of relativistic particles and their possible transitions have not been extensively investigated, except for the transition of a relativistic particle propagation from the ballistic to diffusive regimes in the context of cosmological and ultra high energy particles \cite{0004-637X-612-2-900,PROP:PROP2190390704,PhysRevD.92.083003}. It is, therefore, our purposes for this research to investigate the propagation nature of the relativistic particle governed by the Dirac equation, especially one under random mass and potential fluctuation. Different regimes of mass, energy and fluctuation strength will be studied. 

This article is organized into six sections. In Section 2, we formulate the one-dimensional Dirac particle wave function in terms of the Fourier transforms of eigenspinors. In Section 3, we analytically investigate the eigenspinors' behaviors in some regimes of interests, and we obtain the following results of (i) when using the ultra-relativistic approximation, (ii) when the initial Gaussian wave packet has large spread, and (iii) the dispersion rate as a function of the particle mass. In Section 4, we perform numerical simulations on the free one-dimensional Dirac equation using the Chebyshev expansion method. Our approach yields fast convergent series, and numerical results consistent with analytical results and those obtained by other methods available in literature. Having already bench-marked the numerical algorithm, we direct our attention to study the propagation of a Gaussian wave packet of a Dirac particle with random mass and potential, where the results are presented in Section 5. The discussion of results and their relations to relevant problems and topics in physics are focused in Section 6, and followed by the conclusions.  

\section{Formulation and Research Methodology}
The Dirac equation in $1+1$ dimensions with $\hbar=c=1$ can be written as
\begin{equation}
i\partial_t\Psi(x,t) = \mathbf{H}\Psi(x,t),
\end{equation}
where $\Psi(x,t)$ is a two-component spinor wave function; or, equivalently, in a matrix representation 
\begin{equation}
i\partial_t\underset{\mathlarger{\Psi(x,t)}}{\underbrace{\left(
\begin{array}{ccc}
\Sigma(x,t)\\ X(x,t)
\end{array}
\right)}}=\underset{\mathlarger{\mathbf{H}}}{\underbrace{\left(
\begin{array}{ccc}
V(x) + m & -i\partial_x\\
-i\partial_x & V(x)-m
\end{array}\right)}}
\underset{\mathlarger{\Psi(x,t)}}{\underbrace{\left(
\begin{array}{ccc}
\Sigma(x,t)\\ X(x,t)
\end{array}
\right)}},
\label{FreeDirac}
\end{equation}
where $\mathbf{H}$ is the full Hamiltonian. For the free Dirac equation, i.e. $V(x) = 0$, the Hamiltonian will be denoted by $\mathbf{H}_0$, and the analytical solutions can be easily obtained and are well-known from many methods in literature \cite{Bracken3633,JMathPhys,ReedSimon}:
\begin{eqnarray}
\psi_k(x) &:= &\frac{1}{\sqrt{4\pi\omega(\omega+ m)}}\left(
\begin{array}{c}
+\omega+m\\k
\end{array}
\right)e^{ikx},\\
\phi_k(x) &:=& \frac{1}{\sqrt{4\pi\omega(\omega- m)}}\left(
\begin{array}{c}
-\omega+m\\k
\end{array}
\right)e^{ikx}.
\end{eqnarray}
These choices of solutions, which we use as bases for our future calculations, are inherited with the orthonormality properties
\begin{eqnarray}
\int_{-\infty}^{+\infty}\psi_k(x)^\dagger\psi_{k'}(x){\rm d}x=\int_{-\infty}^{+\infty}\phi^\dagger_k(x)\phi_{k'}(x){\rm d}x &=& \delta(k-k');\\
\int_{-\infty}^{+\infty}\psi_k(x)^\dagger\phi_{k'}(x){\rm d}x=\int_{-\infty}^{+\infty}\phi^\dagger_k(x)\psi_{k'}(x){\rm d}x&=& 0.
\end{eqnarray}
Therefore, the general time-independent solutions to the free Dirac equation can be expressed as
\begin{equation}
\Psi(x) = \int_{-\infty}^{+\infty}\left[\Pi^+(k)\psi_k(x)+\Pi^-(k)\phi_k(x)\right]{\rm d}k\label{eq:Expansion}
\end{equation}
where the Fourier coefficients, $\Pi^+(k)$ and $\Pi^-(k)$ are respectively given by
\begin{eqnarray}
\Pi^+(k) &=& \int_{-\infty}^{+\infty}\psi_k^\dagger(x)\Psi(x){\rm d}x\label{eq:Pi+},\\
\Pi^-(k) &=& \int_{-\infty}^{+\infty}\phi_k^\dagger(x)\Psi(x){\rm d}x.\label{eq:Pi-}
\end{eqnarray}
Consequently, the time evolution of the initial wave function $\Psi(x,0):= \Psi(x)$ is given by $\Psi(x,t) = {\rm e}^{-it\mathbf{H}_0}\Psi(x,0) := {\rm e}^{-it\mathbf{H}_0}\Psi(x)$, or equivalently,
\begin{equation}\label{eq:propagation} 
\Psi(x,t)=\int_{-\infty}^{+\infty}\left[\Pi^+(k)\psi_k(x)e^{i\omega t}+\Pi^-(k)\phi_k(x)e^{-i\omega t}\right]\, dk.
\end{equation}
Unlike in non-relativistic quantum mechanics, the last integral turns out to be very difficult to solve without applying suitable approximations. In the following sections, we investigate the dynamics of a Gaussian-like wave packet using both analytical and numerical methods. For the free Dirac equation, the Fourier transform of the propagator ${\rm e}^{-it\mathbf{H}_0}$ exists, hence the spectral theory allows us to use Eq. \ref{eq:propagation} for approximations in various regimes of interests (Section 3). But when the potential operator is included in the Hamiltonian, it is no longer convenient to express the time evolution in terms of the Fourier coefficients as the propagator ${\rm e}^{-it(\mathbf{H}_0 + \mathbf{V})}$ does not have simple Fourier transform for the operators $\mathbf{H}_0$ and $\mathbf{V}$ need not commute. For the latter case, we resort to numerical methods using the Chebyshev polynomials expansion (Sections 4-5).

\section{Analytical Calculation} 
Consider an initial spinor
\begin{equation}
\Psi(x):=\mathcal{N}_g\left(
\begin{array}{c}
\Sigma_0 \exp(ik_1 x)\\ X_0 \exp(ik_2 x)
\end{array}
\right)\exp{\left(-\frac{x^2}{4\sigma^2}\right)}\label{eq:Gaussian spinor}
\end{equation}
with $\mathcal{N}_g = (2\pi\sigma^2)^{-1/4}$ and $|\Sigma_0|^2 +|X_0|^2 = 1$. Here, $k_1$ and $k_2$ are momentum numbers which can be positive or negative. From Eqs.\ref{eq:Expansion}--\ref{eq:Pi-}, we can expand this spinor in terms of $\Pi^{\pm}(k)$ such that
\begin{eqnarray}
\Pi^+&=\mathcal{N}_g\sigma\left(\frac{(m+\omega)\Sigma_0}{\sqrt{\omega(\omega+m)}}e^{-(k-k_1)^2\sigma^2}+\frac{kX_0}{\sqrt{\omega(\omega+m)}}e^{-(k-k_2)^2\sigma^2}\right);\\
\Pi^-&=\mathcal{N}_g\sigma\left(\frac{(m-\omega)\Sigma_0}{\sqrt{\omega(\omega-m)}}e^{-(k-k_1)^2\sigma^2}+\frac{kX_0}{\sqrt{\omega(\omega-m)}}e^{-(k-k_2)^2\sigma^2}\right).
\end{eqnarray}
Hence, the time development of $\Psi(x)$ is
\begin{equation}
\Psi(x,t) =\frac{\mathcal{N}_g\sigma}{\sqrt{4\pi}}\int_{-\infty}^{+\infty}{{\rm d}k\left\{
\begin{array}{c}
\Sigma_0\!\!\left(
\begin{array}{c}
\frac{\omega+m}{\omega}\\\frac{k}{\omega}
\end{array}
\right)\!e^{-i\omega t+ikx-\sigma^2(k-k_1)^2}
\\+\Sigma_0\!\!\left(
\begin{array}{c}
\frac{\omega-m}{\omega}\\-\frac{k}{\omega}
\end{array}
\right)\!e^{i\omega t+ikx-\sigma^2(k-k_1)^2}\\
+X_0\!\!\left(
\begin{array}{c}
\frac{k}{\omega}\\\frac{\omega-m}{\omega}
\end{array}
\right)\!e^{-i\omega t+ikx-\sigma^2(k-k_2)^2}
\\+X_0\!\!\left(
\begin{array}{ccc}
-\frac{k}{\omega}\\\frac{\omega+m}{\omega}
\end{array}
\right)\!e^{i\omega t+ikx-\sigma^2(k-k_2)^2}
\end{array}
\right\}}\label{eq:analytic_psi}
\end{equation}

\subsection{Ultra-Relativistic Approximation}
In case of $k_1=k_2:=k_0$, e.g. for a particle where the positive- and negative-component spatial momenta have the same direction and magnitude with small mass $m$ (e.g. for neutrinos or photons), we can approximate $\displaystyle\frac{m}{\omega}\approx 0$, and $\omega\approx k.$
Then
\begin{equation}
\Psi(x,t) =\frac{\mathcal{N}_g\sigma}{\sqrt{4\pi}}\int_{-\infty}^{+\infty}{{\rm d}k\left\{
\begin{array}{c}
\Sigma_0\left(
\begin{array}{c}
+1\\+1
\end{array}
\right)e^{ik(x-t)-\sigma^2(k-k_1)^2}
\\+\Sigma_0\left(
\begin{array}{c}
+1\\-1
\end{array}
\right)e^{ik(x+t)-\sigma^2(k-k_1)^2}\\
+X_0\left(
\begin{array}{c}
+1\\+1
\end{array}
\right)e^{ik(x-t)-\sigma^2(k-k_2)^2}
\\+X_0\left(
\begin{array}{c}
-1\\+1
\end{array}
\right)e^{ik(x+t)-\sigma^2(k-k_2)^2}
\end{array}
\right\}}
\end{equation}
The probability density $P(x,t)=\Psi^\dagger(x,t)\Psi(x,t)$ is then
\begin{equation}\label{eq:probdensity}
P(x,t) =\frac{\mathcal{N}_g^2}{2}(\Sigma_0-X_0)^2e^{-\frac{(x+t)^2}{2\sigma^2}}+\frac{\mathcal{N}_g^2}{2}(\Sigma_0+X_0)^2e^{-\frac{(x-t)^2}{2\sigma^2}}
\end{equation}
which are two Gaussian wave packets moving with the speed of light to the left and to the right, as expected in the limit $m\rightarrow0$.

\subsection{Large-$\sigma$ Approximation}
For $\sigma$ sufficiently large, the Gaussian function in the integrand of Eq. \ref{eq:analytic_psi} will vanish very quickly for $|k-k_j|>1/(2\sigma)$, with $k_j$ being either $k_1$ or $k_2$. Thus, we can perform Taylor expansion around $k = k_j$ to the second order.  Denoting $\omega(k_j):=\omega_j$, we obtain the approximations
\begin{eqnarray}
&& \frac{k}{\omega} \approx \underset{A_j}{\underbrace{\frac{k_j}{\omega_j}}}+\underset{B_j}{\underbrace{\frac{m^2}{\omega_j^3}}}(k-k_j)\underset{C_j}{\underbrace{-\frac{3}{2}\frac{k_jm^2}{\omega_j^5}}}(k-k_j)^2\\
&&\frac{\omega\pm m}{\omega} \approx \underset{D^\pm_j}{\underbrace{\frac{\omega_j\pm m}{\omega_j}}}\underset{E^\pm_j}{\underbrace{\mp\frac{k_jm}{\omega_j^3}}}(k-k_j)\underset{F^\pm_j}{\underbrace{\mp\frac{1}{2}\frac{m(m^2-2k_j^2)}{\omega_j^5}}}(k-k_j)^2
\end{eqnarray}
\newpage
\begin{eqnarray}
&& i(kx\pm\omega t)- \sigma^2(k-k_j)^2\nonumber \approx  i\underset{\varphi^\pm_j}{\underbrace{(k_jx\pm\omega_jt)}}+i\underset{\xi^\pm_j}{\underbrace{\frac{\omega_jx\pm k_jt}{\omega_j}}}(k-k_j)\\&&\hspace*{5cm}-\underset{(\sigma^\pm_j)^2}{\underbrace{\left(\sigma^2\mp i\frac{m^2t}{2\omega_j^3}\right)}}(k-k_j)^2
\end{eqnarray}
Then our approximated solution becomes
\begin{equation}
\!\!\!\!\!\!\!\!\!\!\!\!\!\!\!\!\!\!\!\!\!\!\!\!\!\!\!\!\!\!\!\!\!\!\!\!\!\!\!\!\!\!\!\!\!\!\!\!
\Psi(x,t) =\frac{\mathcal{N}_g\sigma}{\sqrt{4\pi}}\int_{-\infty}^{+\infty}{{\rm d}k\left[
\begin{array}{ccc}
\Sigma_0\left(
\begin{array}{ccc}
D^+_1-E^+_1(k-k_1)-F^+_1(k-k_1)^2\\
A_1+B_1(k-k_1)+C_1(k-k_1)^2
\end{array}
\right)e^{i\varphi^-_1+i\xi^-_1(k-k_1)-(\sigma^-_1)^2(k-k_1)^2}\\
+\Sigma_0\left(
\begin{array}{ccc}
D^-_1-E^-_1(k-k_1)-F^-_1(k-k_1)^2\\
-A_1-B_1(k-k_1)-C_1(k-k_1)^2
\end{array}
\right)e^{i\varphi^+_1+i\xi^+_1(k-k_1)-(\sigma^+_1)^2(k-k_1)^2}\\
+X_0\left(
\begin{array}{ccc}
A_2+B_2(k-k_j)+C_2(k-k_2)^2\\
D_2^-+E_2^-(k-k_j)+F_2^-(k-k_2)^2
\end{array}
\right)e^{i\varphi^-_2+i\xi^-_2(k-k_2)-(\sigma^-_2)^2(k-k_2)^2}\\
+X_0\left(
\begin{array}{ccc}
-A_2-B_2(k-k_j)-C_2(k-k_2)^2\\
D_2^++E_2^+(k-k_j)+F_2^+(k-k_2)^2
\end{array}
\right)e^{i\varphi^+_2+i\xi^+_2(k-k_2)-(\sigma^+_2)^2(k-k_2)^2}
\end{array}
\right]}
\label{largessol}
\end{equation}
The integral can be solved explicitly, but we will leave it in this form. For large-$\sigma$ approximation, we consider the second-order approximation around $k=k_j$
\begin{equation}
f_{\rm apx}(k)=f(k_j)+f'(k_j)(k-k_j)+\frac{1}{2}f''(k_j)(k-k_j)^2,
\end{equation}
and determine how large $\sigma$ is possible. To that end, the relative errors at $|k-k_j|=n\sigma_k$ should be very small, given by
\begin{equation}
\frac{\delta f(n\sigma_k)}{f(n\sigma_k)} \approx \frac{1}{6}\frac{f^{(3)}(k_j)}{f(n\sigma_k)}(n\sigma_k)^3=\epsilon.
\label{solvefor}
\end{equation}
where $n$ and $\epsilon$ are at our disposal, but $n \sim 5$ and $\epsilon \sim 10^{-3}$ are chosen for numerical calculations. Let $\sigma_{f;k_j}$ be a solution to Eq.\ref{solvefor} which corresponds to the function $f$ at point $k_j$, then our appropriated $\sigma$ can be taken as
\begin{equation}
\sigma=\max\left\{
\sigma_{\frac{k}{\omega};k_1},\sigma_{\frac{k}{\omega};k_2},\sigma_{\frac{\omega\pm m}{\omega};k_1},
\sigma_{\frac{\omega\pm m}{\omega};k_2}
\right\}
\end{equation}
This value of $\sigma$ gives minimal error to the Taylor expansions, which we will use in all other analytical calculations.
\subsection{Dispersion Rate}
From Eq.\ref{eq:probdensity}, the wave propagates symmetrically to the left and to the right, simultaneously. In some cases, the divergence of left-moving and right-moving waves can be wrongly interpreted as a dispersion, while in fact they are not. So, in this section, we will consider the dispersion in a Gaussian function.
Consider Eq.\ref{largessol}, where the integrand of the exponential factor can be rearranged as follows:
\begin{eqnarray}
-\big(\sigma_j^{\pm}\big)^2(k-k_j)^2+ i(k-k_j)\xi_j^\pm+ i\varphi_j^\pm \nonumber\\
\;\;\;\;\;\;\;\;\;\;\;\;=-\big(\sigma_j^{\pm}\big)^2\left(k-\!k_j-i\frac{\xi^\pm_j}{2\big(\sigma_j^{\pm}\big)^2}\right)^2 + i\varphi^\pm_j-\frac{\big(\xi_j^{\pm}\big)^2}{4\big(\sigma_j^{\pm}\big)^2}
\end{eqnarray}
Integrating over $k$ and considering the non-Gaussian parts as constants, since they would not give rise to the dispersion behavior, the solution in Eq.\ref{largessol} can be written as
\begin{equation}
\Psi(x,t)=\mathlarger{\sum}_{j=1,2}\left[
\left(
\begin{array}{c}
C_{1j}\\C_{2j}
\end{array}
\right)\exp\left(-\frac{1}{4}\Big(\frac{\xi_j^{+}}{\sigma_j^{+}}\Big)^2\right)+\left(
\begin{array}{c}
C_{3j}\\C_{4j}
\end{array}
\right)\exp\left(-\frac{1}{4}\Big(\frac{\xi_j^{-}}{\sigma_j^{-}}\Big)^2\right)\right].
\end{equation}
As a result, the probability density (up to the Gaussian part) is
\begin{equation}
P(x,t)=\sum_{j=1,2}\left[D_j\exp\left(-\frac{(x-k_jt/\omega_j)^2}{2\sigma^2_j(t)}\right)+E_j\exp\left(-\frac{(x+k_jt/\omega_j)^2}{2\sigma^2_j(t)}\right)\right]
\end{equation}
with
\begin{equation}
\sigma^2_j(t)=\sigma^2+\frac{m^4t^2}{(m^2+k_j^2)^3\sigma^2}.
\label{width_dispersion}
\end{equation}
The wave packet disperses with the function $\sigma_j(t)$ that depends on the particle's mass $m$, where the \emph{dispersion rate} $R(m)$ for large $t$ is
\begin{equation}
R(m) = \lim_{t \rightarrow \infty}\dot{\sigma}_j(t) 
=\frac{m^2}{\sigma(m^2+k_j^2)^{3/2}}\sim\left\{
\begin{array}{cc}
\displaystyle\frac{m^2}{\sigma k_j^3},\, &{\rm for\;small\;mass\;}m\\
\displaystyle\frac{1}{\sigma m},\, &{\rm for\;large\;mass\;}m
\end{array}
\right..
\end{equation}
Interestingly, the relation suggests that both the super-massive and massless particles will not disperse in free space.

It should be remarked that for large $\sigma$ and in the ultra-relativistic limit, 
the approximation works poorly as $k_0\rightarrow0$ and $m\rightarrow0$, because the approximation of $\displaystyle \frac{k}{\omega}$ gives relatively large errors. On the other hand, for large $m$, the approximation works very well for every $k_0$. Thus, the only region that the approximation goes wrong is where both $k_0$ and $m$ approach zero simultaneously.

\section{Numerical Calculation}
From Eq.\ref{FreeDirac}, the Dirac equation in 1+1 dimensions under the time-independent scalar potential $V(x)$ can be written as 
\begin{equation}
i\partial_t\underset{\mathlarger{\Psi(x,t)}}{\underbrace{\left(
\begin{array}{c}
\Sigma(x,t)\\ X(x,t)
\end{array}
\right)}}=\underset{\mathlarger{\mathbf{H}}}{\underbrace{\left(
\begin{array}{cc}
V(x)+m & -i\partial_x\\
-i\partial_x&V(x)-m
\end{array}
\right)}}\underset{\mathlarger{\Psi(x,t)}}{\underbrace{\left(
\begin{array}{c}
\Sigma(x,t)\\ X(x,t)
\end{array}
\right)}}.
\label{InteractingDirac}
\end{equation}
In this case, the Fourier transform method has its limitation since the propagation has compound effects of the potential and kinetic energy, which cannot be separated from one another because their operators do not commute. Various numerical procedures have been employed to investigate the propagator $\mathcal{U}(t):=\exp\left(-it\mathbf{H}\right)$; see \cite{RevModPhys.69.731,Kalman,PhysRevA.57.3554} and references therein. We have chosen to expand $\mathcal{U}(t)$ in the Chebyshev polynomials.

\subsection{Chebyshev Polynomial Expansion}
The Chebyshev polynomials of the first kind $T_k(x)$ are defined recursively by \cite{Kalman}
\begin{eqnarray}
T_0(x) &=& 1,\nonumber\\
T_1(x) &=& x,\nonumber\\
T_{k+1}(x) &=& 2xT_k(x)-T_{k-1}(x);\;\;\;\;\;|x|<1, 
\end{eqnarray}
with a special property that
\begin{equation}
T_k(\cos\theta)=\cos(k\theta).
\end{equation}
These polynomials are the solutions to the Sturm-Liouville differential equation
\begin{equation}
(1-x^2)y''-xy'+k^2y = 0.
\end{equation}
Hence, their orthogonal relation is given by
\begin{equation}
\int_{-1}^{+1}\frac{T_k(x)T_{k'}(x)}{\sqrt{1-x^2}}\;dx=\frac{\pi}{2}\delta_{kk'}(1+\delta_{k0}).
\end{equation}
It is well known that we can expand an exponential function in terms of $T_k(x)$ as
\begin{equation}
e^{-iax} = \sum_{k=0}^{\infty} A_k T_k(x),
\end{equation}
where $A_k$ can be computed in terms of the Bessel function of the first kind $J_k$, and the second kind $I_k$ by
\begin{eqnarray*}
A_k&=&\Big(\frac{2-\delta_{k0}}{\pi}\Big) \int_{-1}^{+1}\frac{T_k(x)e^{-iax}}{\sqrt{1-x^2}}dx\\
&=&\Big(\frac{2-\delta_{k0}}{\pi}\Big) \int_\pi^0\frac{T_k(\cos\theta)e^{-ia\cos\theta}}{\sin\theta}d\cos\theta\\
&=&\Big(\frac{2-\delta_{k0}}{\pi}\Big) \int_0^\pi\cos(k\theta)e^{-ia\cos\theta}d\theta\\
&=&(2-\delta_{k0})I_k(-ia) = (2-\delta_{k0})(-i)^kJ_k(a)
\end{eqnarray*}
One can see that the Chebyshev expansion converges rapidly because the coefficients $A_k$ decay exponentially if $k > a$. By choosing the parameter $a$ suitably, we can expand the exponential operator up to the second-degree polynomials. In our case, we can set the upper and the lower limits of the Hamiltonian as
\begin{eqnarray}
E_{\rm max} &:=& V_{\rm max}+\sqrt{p_{\rm max}^2+m^2}\\
E_{\rm min} &:=& V_{\rm min}-\sqrt{p_{\rm max}^2+m^2}\end{eqnarray}
In the lower limit, we use maximum momentum under the square-root since the energy is lowest in the case of high-momentum negative-energy solution. We can approximate this by considering the maximum momentum of the non-relativistic particle-in-a-box with lattice constant $d_x$:
\begin{equation}
p_{\rm x}:=\frac{\pi N}{L}=\frac{\pi}{d_x}
\end{equation}
Thus, the Chebyshev interval is
\begin{eqnarray}
E_{\rm max} &:=& V_{\rm max}+\sqrt{\frac{\pi^2}{d_x^2}+m^2}\\
E_{\rm min} &:=& V_{\rm min}-\sqrt{\frac{\pi^2}{d_x^2}+m^2}.
\end{eqnarray}
Now let us define
\begin{equation}
\varepsilon :=\frac{2\mathbf{H}-(E_{\rm x}+E_{\rm n})\mathbf{I}}{E_{\rm x}-E_{\rm n}},
\end{equation}
so that eigenvalues of $\varepsilon$ are inclusively between $1$ and $-1$. We then obtain
\begin{equation}
\mathbf{H} = \mathbf{I}\Big(\frac{E_{\rm x}+E_{\rm n}}{2}\Big)+\varepsilon\Big(\frac{E_{\rm x}-E_{\rm n}}{2}\Big).
\end{equation}
Then the time evolution operator to the second-order expansion is
\begin{eqnarray}
\mathcal{U}(t) &\approx& e^{-it\frac{E_{\rm x}+E_{\rm n}}{2}}\left[J_0T_0(\varepsilon)-2iJ_1T_1(\varepsilon)-2J_2T_2(\varepsilon)\right]\nonumber\\
&=& e^{-it\frac{E_{\rm x}+E_{\rm n}}{2}}\left[\left(J_0+2J_2\right)\mathbf{I}+\left(-2iJ_1\right)\varepsilon+\left(-4J_2\right)\varepsilon^2\right]
\end{eqnarray}
where $J_k = J_k\left(t\frac{E_{\rm x}-E_{\rm n}}{2}\right)$ for $k = 0,\; 1,\; {\rm or}\;2$. Here, we can adjust $t$ for  better precision.

\subsection{Numerical Method Implementation}
The solution to the Dirac equation $\psi(x,t)$ can be written in a numerical procedure as
\begin{equation}
\psi(x,t)=\left(
\begin{array}{c}
\sigma(x,t)\\ \chi(x,t)
\end{array}
\right)
\longrightarrow \psi(t):=\left(
\begin{array}{c}
\sigma_0(t)\\
\vdots\\
\sigma_{N-1}(t)\\
\chi_0(t)\\
\vdots\\
\chi_{N-1}(t)
\end{array}
\right)=\left(
\begin{array}{c}
\psi_0(t)\\
\psi_1(t)\\
\vdots\\
\vdots\\
\psi_{2N-2}(t)\\
\psi_{2N-1}(t)
\end{array}
\right).
\end{equation}
With this representation, we can derive the differential operator as follows. Consider the symmetric differentiation \cite{CrankNicolson}
\begin{equation}
\partial_xy=f(x)\longrightarrow \frac{y_{j+1}-y_{j-1}}{2d_x}=\frac{f_{j+1}+2f_j+f_{j-1}}{4}
\end{equation}
where $d_x$ is the spatial difference set to be constant along the space and time. We can write this in the matrix form as
\begin{eqnarray*}
\frac{1}{2d_x}\left(
\begin{array}{ccccc}
0&1&&&\\
-1&0&1&&\\
&-1&\ddots&&\\
&&&0&1\\
&&&-1&0\\
\end{array}
\right)
\left(
\begin{array}{c}
y_0\\
y_1\\
\vdots\\
y_{N-2}\\
y_{N-1}
\end{array}
\right)
=\frac{1}{4}\left(
\begin{array}{ccccc}
2&1&&&\\
1&2&1&&\\
&1&\ddots&&\\
&&&2&1\\
&&&1&2\\
\end{array}
\right)
\left(
\begin{array}{ccc}
f_0\\
f_1\\
\vdots\\
f_{N-2}\\
f_{N-1}
\end{array}
\right)
\end{eqnarray*}
where we have applied the boundary conditions $y_{-1}=y_{N}=f_{-1}=f_N=0$, which do not affect the system if $y_j$ is localized far from the boundary. To get the differential matrix operator, we have to multiply the equation from the left by the inverse of the square matrix on the right hand side:
\begin{equation}
\mathcal{D} :=\frac{2}{d_x}\left(
\begin{array}{ccccc}
2&1&&&\\
1&2&1&&\\
&1&\ddots&&\\
&&&2&1\\
&&&1&2\\
\end{array}
\right)^{-1}\left(
\begin{array}{ccccc}
0&1&&&\\
-1&0&1&&\\
&-1&\ddots&&\\
&&&0&1\\
&&&-1&0\\
\end{array}\right)
\end{equation}
or, equivalently,
\begin{equation}
\mathcal{D}_{ij} = \sum_{k=0}^{N-1}\frac{2}{d_x}\cdot(\delta_{k, j+1}-\delta_{k+1, j})\cdot\left\{
\begin{array}{ccc}
{\frac{(-1)}{N+1}}^{i+k}(N-k)(i+1)&;&i\leq k,\;i+k<N\\
\mathcal{D}_{ki}&;&i>k,\;i+k<N\\
\mathcal{D}_{N-k-1\;N-i-1}&;&{\rm else}
\end{array}
\right\}.
\end{equation}
Using this definition, we can write the Hamiltonian as
\begin{equation}
\mathbf{H}=\left(
\begin{array}{cccccc}
V_0+m&\dots&0&&&\\
\vdots&\ddots&\vdots&&-i\mathcal{D}&\\
0&\dots&V_{N-1}+m&&&\\
&&&V_0-m&\dots&0\\
&-i\mathcal{D}&&\vdots&\ddots&\vdots\\
&&&0&\dots&V_{N-1}-m\\
\end{array}
\right)
\end{equation}

\subsection{Results from Numerical Simulations}
\begin{figure}[H]
\centering
\includegraphics[scale=0.7]{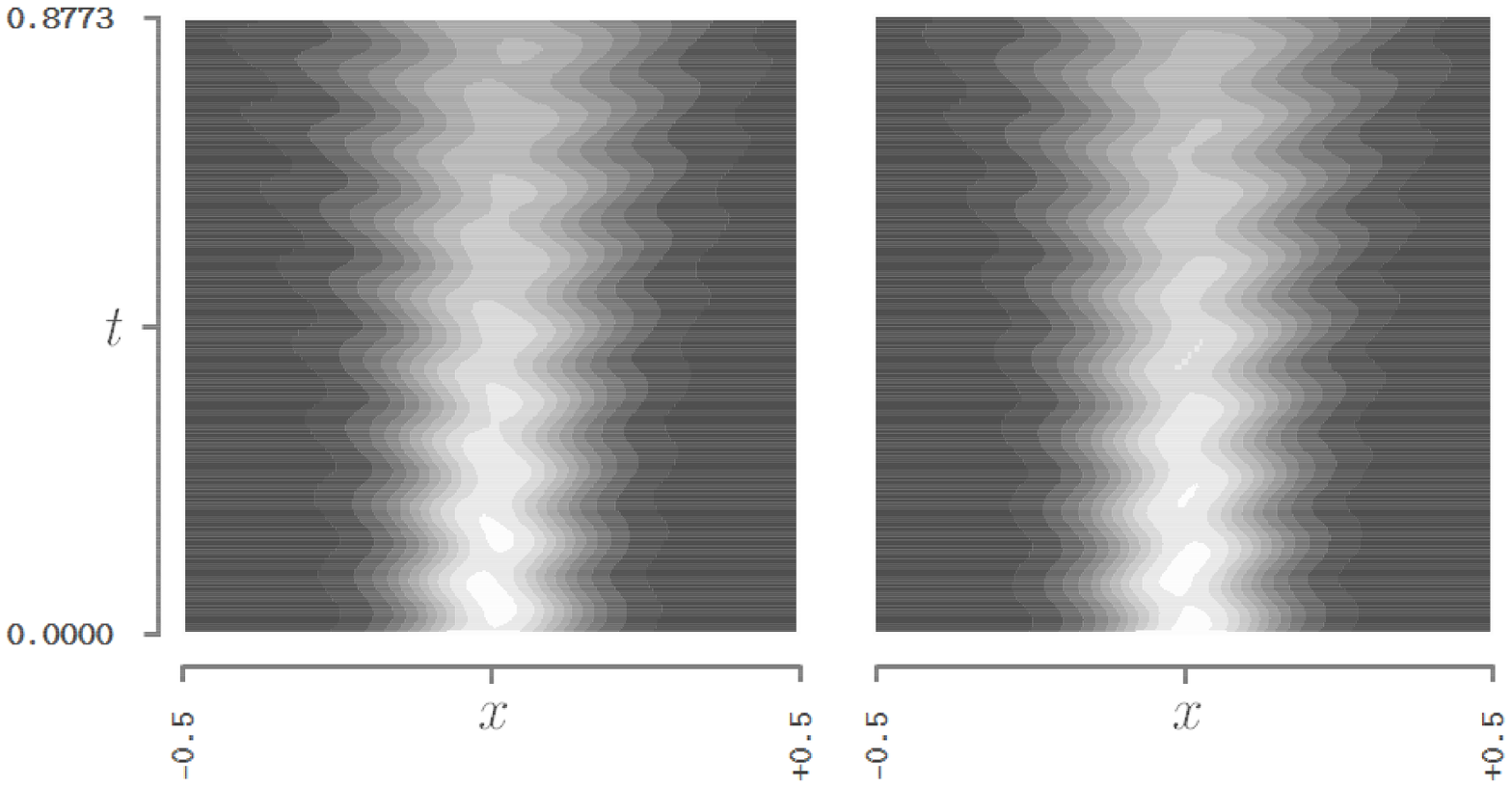}
\caption{(left) Numerical solution of $P(x,t)$ with the initial wave packet given by Eq.\ref{static} with horizontal and vertical axes as space and time (arbitrary units) respectively for $t>0$. (right) analytical solution for the same conditions.}
\label{num0}
\end{figure}

Numerical results on the propagation of a free Dirac particle have been achieved before by many authors \cite{Alhaidari,PhysRevLett.106.256803,Thaller}, which we can use to verify our results along with the analytical solutions in the previous sections. Note that the time-development of wave packets in Dirac equation will be different from those in Schrodinger equation due to the possibility of interference between the two components of the spinor, which is known as the Zitterbewegung, and can be identified as the oscillation of expectation value of position $\langle x\rangle$ in some of the solutions.  The first example is for the initial Gaussian wave packet with zero momentum:
\begin{equation}
\Psi(x,0)=\mathcal{N}\exp\left(-x^2/4\sigma^2\right)\left(
\begin{array}{c}
1\\1
\end{array}
\right)
\label{static}
\end{equation}
with $\sigma=0.1$ and $m=30$. Fig.\ref{num0} shows the comparison between our numerical and analytical results, which agree very well with those obtained by Ref. \cite{Thaller}.

\begin{figure}[H]
\centering
\includegraphics[scale=0.7]{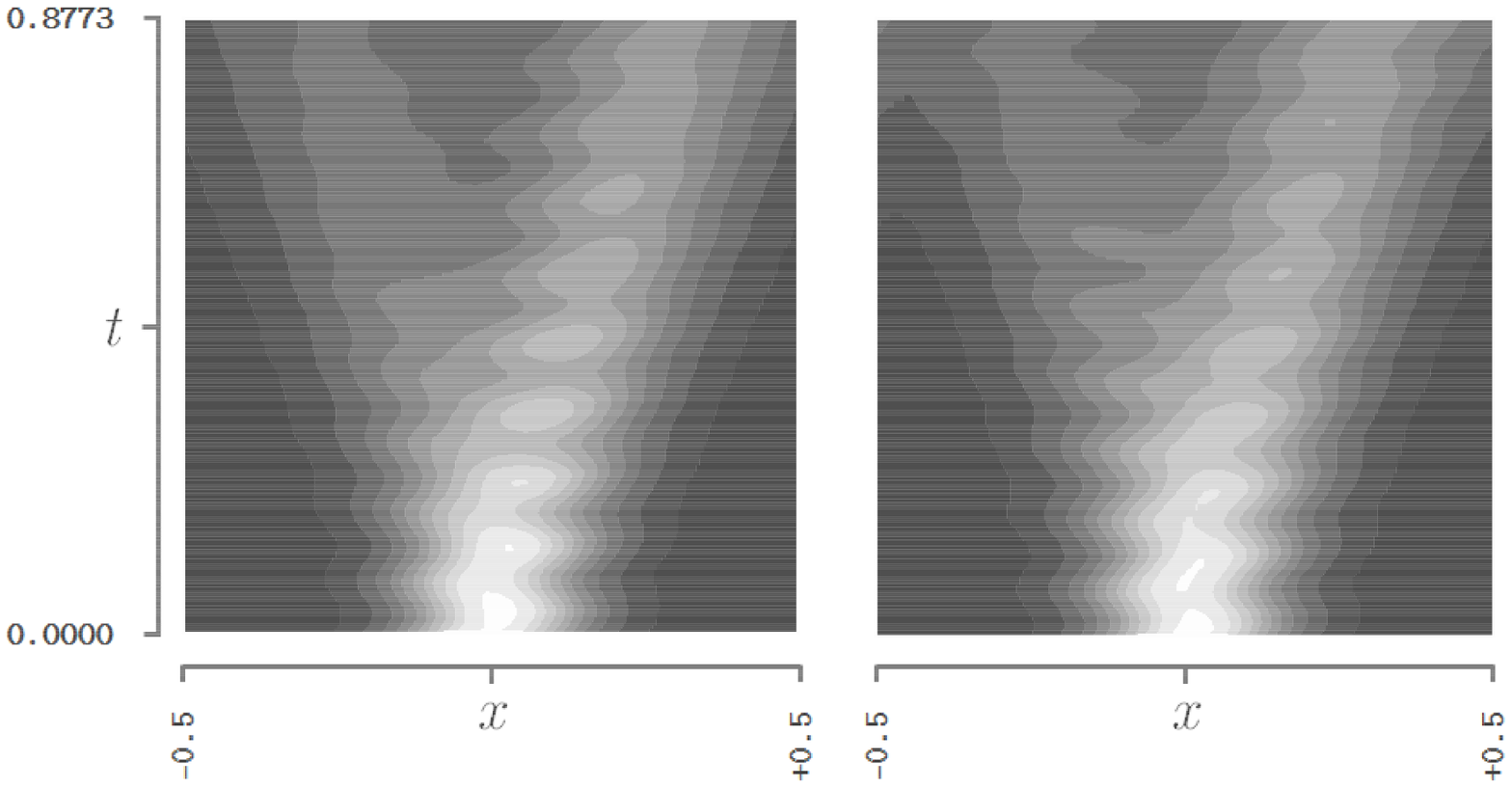}
\caption{(left) Numerical solution of $P(x,t)$ with the initial wave packet given by Eq.\ref{opposite} for $t>0$. (right) analytical solution for the same conditions.}
\label{num1}
\end{figure}

The second example is for the initial Gaussian wave packet with nonzero momentum
\begin{equation}
\Psi(x,0)=\mathcal{N}\exp\left(-x^2/4\sigma^2+ik_0x\right)\left(
\begin{array}{c}
1\\1
\end{array}
\right)
\label{opposite}
\end{equation}
with $\sigma=0.1$, $m=30$, and $k_0=10$. Fig.\ref{num1} shows the comparison between our numerical and analytical results, which also agree very well with those obtained from Ref.\cite{Thaller}

\begin{figure}[H]
\centering
\includegraphics[scale=0.7]{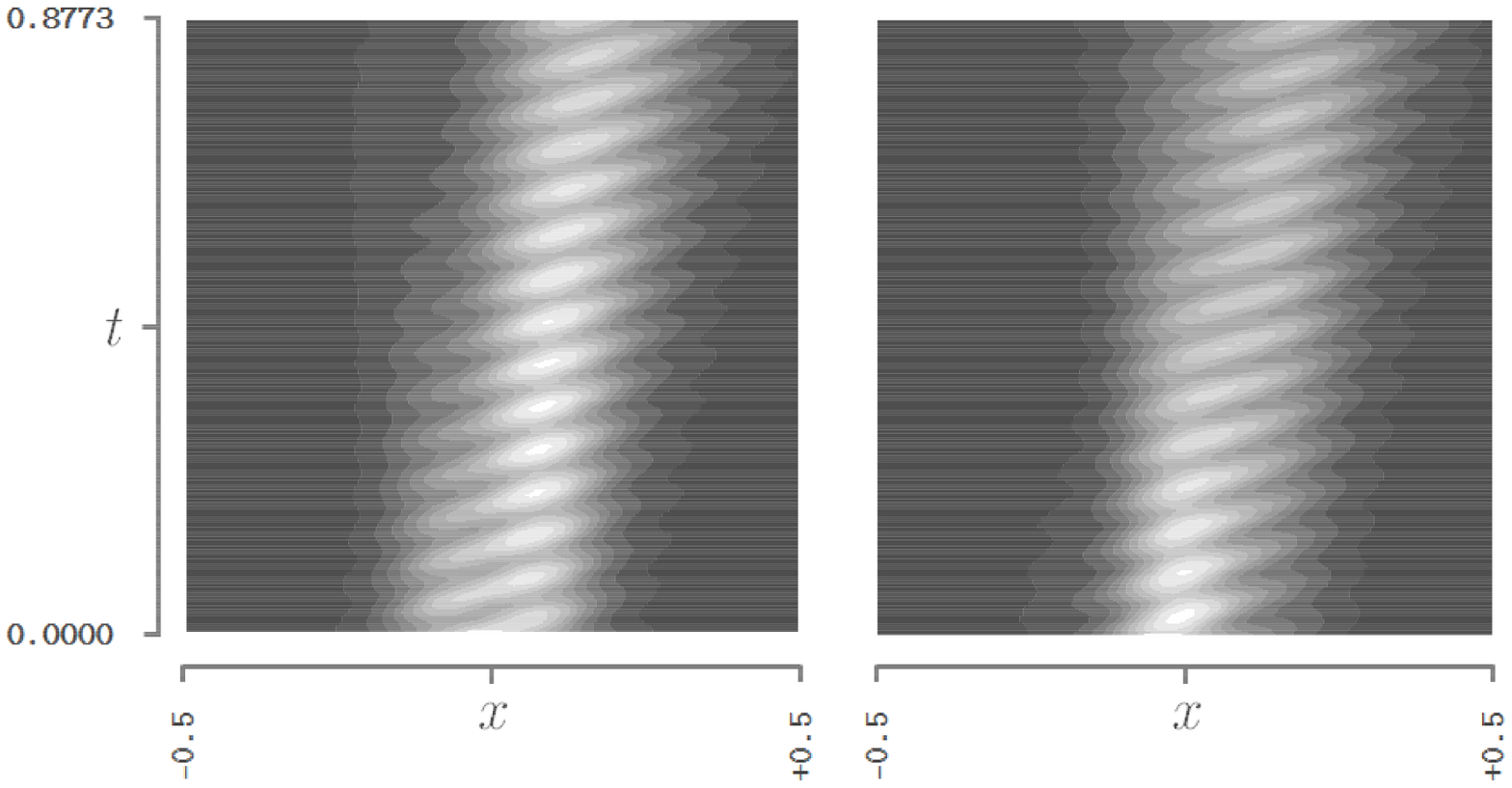}
\caption{(left) Numerical solution of $P(x,t)$ with the initial wave packet given by Eq.\ref{paralel} for $t>0$. (right) analytical solution for the same condition.}
\label{num2}
\end{figure}

The third example is for the initial Gaussian wave packet whose positive- and negative-energy components of the momentum have opposite directions (but with the same spatial direction), 
\begin{equation}
\Psi(x,0)=\mathcal{N}\exp\left(-x^2/4\sigma^2\right)\left(
\begin{array}{c}
\exp(+ik_0x)\\\exp(-ik_0x)
\end{array}
\right)
\label{paralel}
\end{equation}
with $\sigma=0.1$, $m=50$, and $k_0=10$. The result is a constant ripple along the path as shown in Fig.\ref{num2}. Infact, this kind of ripple is not the Zitterbwewgung since an expectation value for position is not oscillating. The only slight difference between our result and those obtained from Ref.\cite{Thaller} stems from the expansion of the packet due to the mass of a particle.
\\[1cm]

\section{Dirac Equation with Random Mass and Potential}
In this section, we investigate the spread of the spinor wave function of a Dirac particle when it is subject to fluctuating potential or a random mass. In case of the random potential, we set the potential to be a random variable with uniform distribution on $[-V_0, V_0]$. And in case of the random mass, we set the mass to be a random variable with uniform distribution on $[m-m_0, m+m_0]$. This does not necessarily mean all potential or mass distributions are of this random type; a more natural one should have Gaussian distribution. For simplicity, we perform simulations with uniformly distributed variables. Theoretically, the distribution of randomness should be separated into two types; one with a probability density and the other having discrete distribution. It is hypothesized that localization should behave differently for different types \cite{Suwanna} but should yield similar behaviors for the same type. For sufficiently large randomness strength $\sigma_r$ (defined by the standard deviation of the distribution), localization should set in. 

To calculate a quantity that can indicate localization of a waveform in one spatial dimension, we define a \textit{localization functional}
\begin{equation}
L[p(x)]:=\int_{-\infty}^{\infty}\left|\sqrt{p(x)}\frac{{\rm d}^2}{{\rm d}x^2}\sqrt{p(x)}\right|{\rm d}x
\label{localization}
\end{equation}
where $p(x) = |\psi(x)|^2$ is the probability density of a particle. $L[p(x)]$ is in fact one of many functionals with the following properties: (i) it  has direct relation to Gaussian localization width; (ii) it has direct relation to the localization length, and (iii) for the summation of two identical normalized peaks, it yields the same value as that of a single peak. In particular, for a standard normalized Gaussian function, it follows that 
\begin{equation}
L\left[\frac{\exp(-x^2/2\sigma^2)}{\sigma\sqrt{2\pi}}\right]=\frac{C}{\sigma^2},
\end{equation}
with a universal constant 
\[C:=-\frac{1}{4}+\frac{1}{2}{\rm erf}(1)+\frac{\exp(-1)}{\sqrt{\pi}}\approx0.3789041452.\]
Then, we can calculate an approximated localization width for $p(x)$ by
\begin{equation}
W[p(x)]:=\sqrt{C/L[p(x)]}.
\label{localization_width}
\end{equation}
It is easy to verify that if we have a convex combination of $N$ identical normalized Gaussian distributions as
\begin{equation}
p(x):=\sum_{j=1}^N\frac{\alpha_j}{\sigma\sqrt{2\pi}}e^{-(x-x_j)^2/2\sigma^2,}
\end{equation}
with $\alpha_j > 0$ and $\sum_j\alpha_j=1$, and if the wave packets have supports far enough from each other, we can approximate the integral in Eq.\ref{localization}
\begin{eqnarray}
L[p(x)]&\approx\sum_j\alpha_jL\left[\frac{\exp(-x^2/2\sigma^2)}{\sigma\sqrt{2\pi}}\right]=\frac{C}{\sigma^2},
\end{eqnarray}
which yields the same value of $W[p(x)]$ as that of the one-packet localization. In general, if we have a summation of packets with different widths, the representation width $W[p(x)]$ can be approximated with the weighted generalized mean
\begin{equation}
W\left[\sum_j \alpha_jp_j(x)\right]\approx\left(\sum_j\frac{\alpha_j}{W^2[p_j(x)]}\right)^{-\frac{1}{2}}.
\end{equation}

For Anderson localization, it can be shown that $W[p(x)]$ is directly proportional to the traditional definition of localization length \cite{Croy2011,Kramer}, where the constant of proportionality is shape-dependent. To show this, let $p(x)$ be a normalized probability density which decays exponentially as $x \rightarrow \pm\infty$. It is worth noting that under the transformation
\begin{equation}
T_\Lambda:p(x)\rightarrow p_\Lambda(x)=\frac{1}{\Lambda}p\left(\frac{x}{\Lambda}\right),
\end{equation}
the transformed function $p_\Lambda(x)$ is still normalized for any finite $\Lambda > 0$, and decays exponentially as $x \rightarrow \pm\infty$ (though with different rates).
Without the loss of generality, we can always redefine $\Lambda$ and such the choice of $\Lambda$ is simply the well-known localization length \cite{Croy2011,Kramer}. Hereinafter, we will call the unscaled ($\Lambda=1$) probability density as $p_1(x)$ for the consistency of notation. Then
\begin{eqnarray}
L[p_\Lambda(x)]&=\int_{-\infty}^{+\infty}\left|\sqrt{p_\Lambda(x)}\frac{{\rm d}^2}{{\rm d}x^2}\sqrt{p_\Lambda(x)}\right|{\rm d}x\\
&=\frac{1}{\Lambda^2}\int_{-\infty/\Lambda}^{+\infty/\Lambda}\left|\sqrt{p_1\left(\frac{x}{\Lambda}\right)}\frac{{\rm d}^2}{{\rm d}\left(\frac{x}{\Lambda}\right)^2}\sqrt{p_1\left(\frac{x}{\Lambda}\right)}\right|{\rm d}\left(\frac{x}{\Lambda}\right)\\
&=\frac{1}{\Lambda^2}L[p_1(x)],
\end{eqnarray}
where we assume that $p_1(x) = |\psi(x)|^2$ is at least twice differentiable. Consequently,
\begin{eqnarray}
W[p_\Lambda(x)]&=\sqrt{\frac{C}{L[p_\Lambda(x)]}}=\Lambda\sqrt{\frac{C}{L[p_1(x)]}},
\end{eqnarray}
so the localization length $\Lambda$ and $W[p(x)]$ are equivalent up to the multiplicative factor $\sqrt{C/L[p_1(x)]}$.

\begin{figure}[H]
\centering
\includegraphics[scale=0.7]{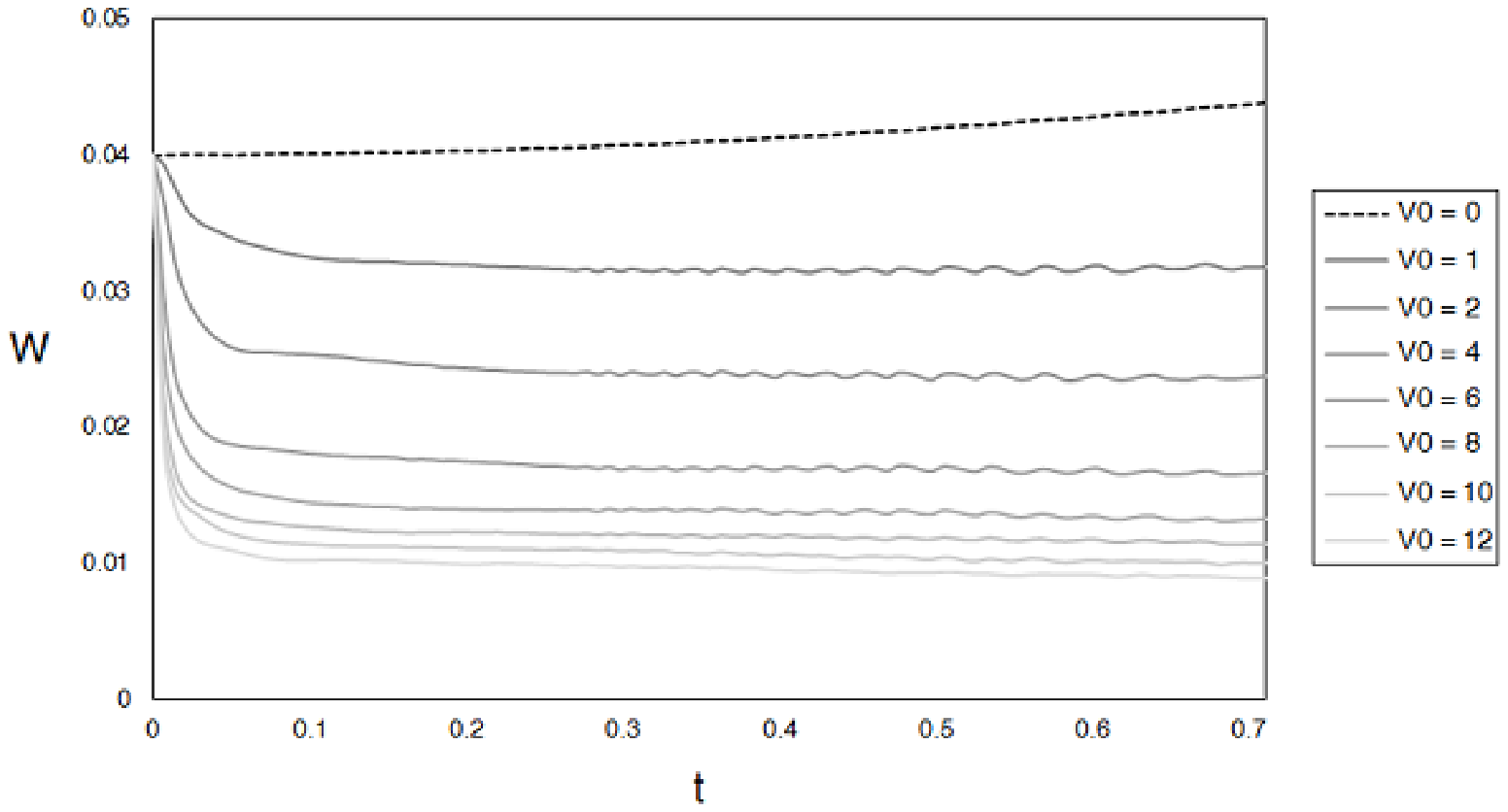}\\
\includegraphics[scale=0.7]{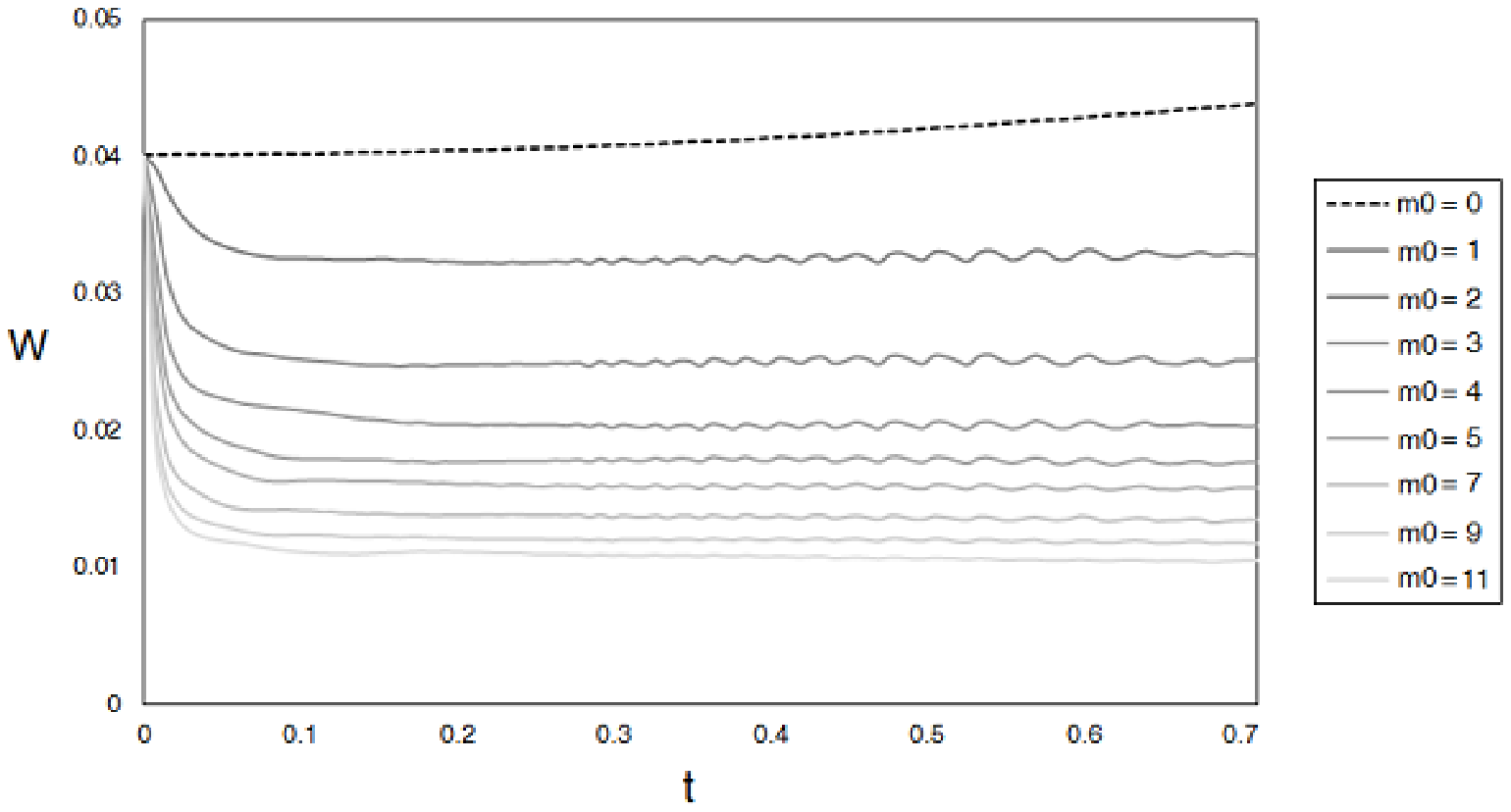}
\caption{Time evolution of localization width $W$ ($0<t<0.71064$) for each random strength $V_0$ and $m_0$}
\label{evol}
\end{figure}

\begin{figure}[H]
\centering
\includegraphics[scale=0.57]{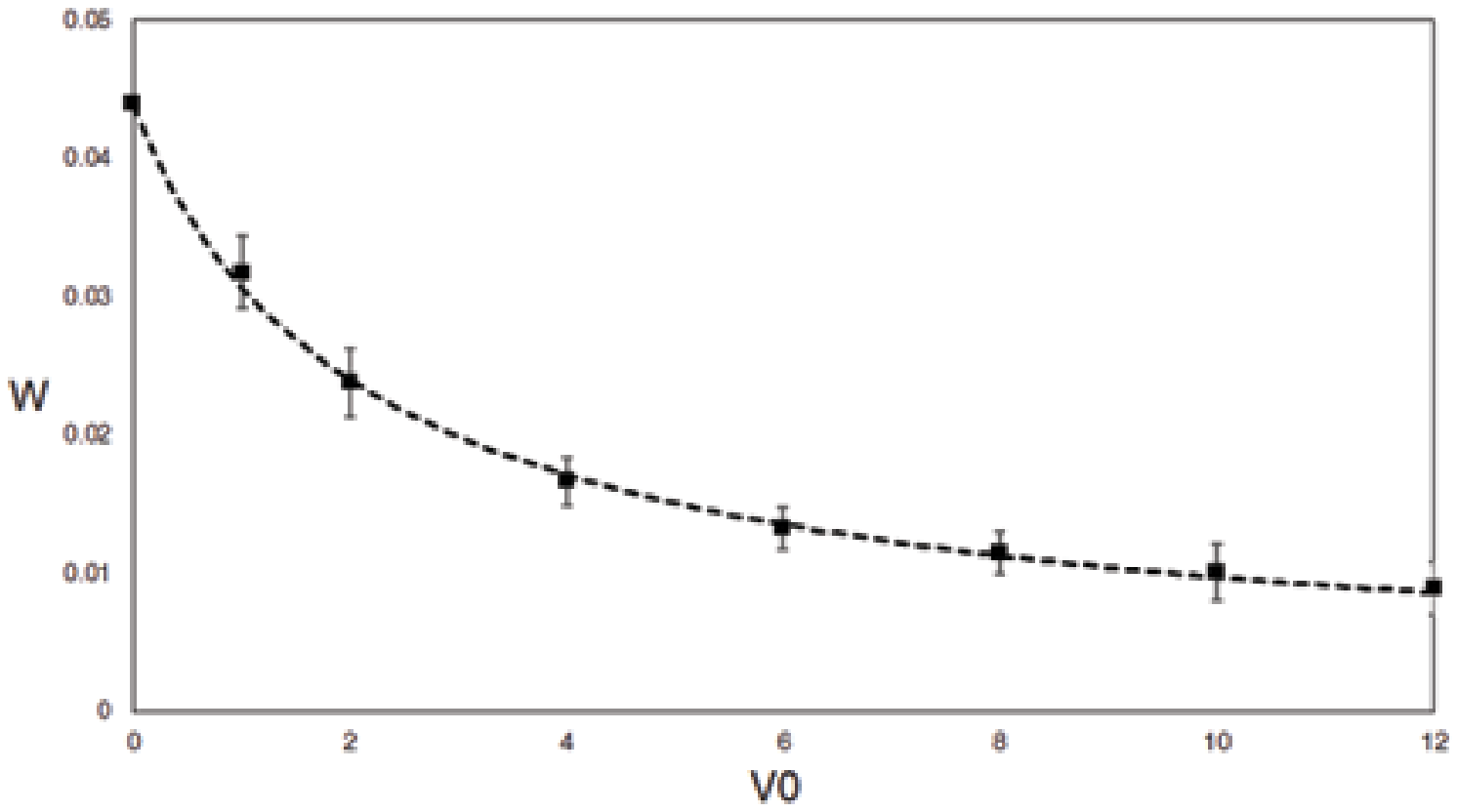}
\includegraphics[scale=0.57]{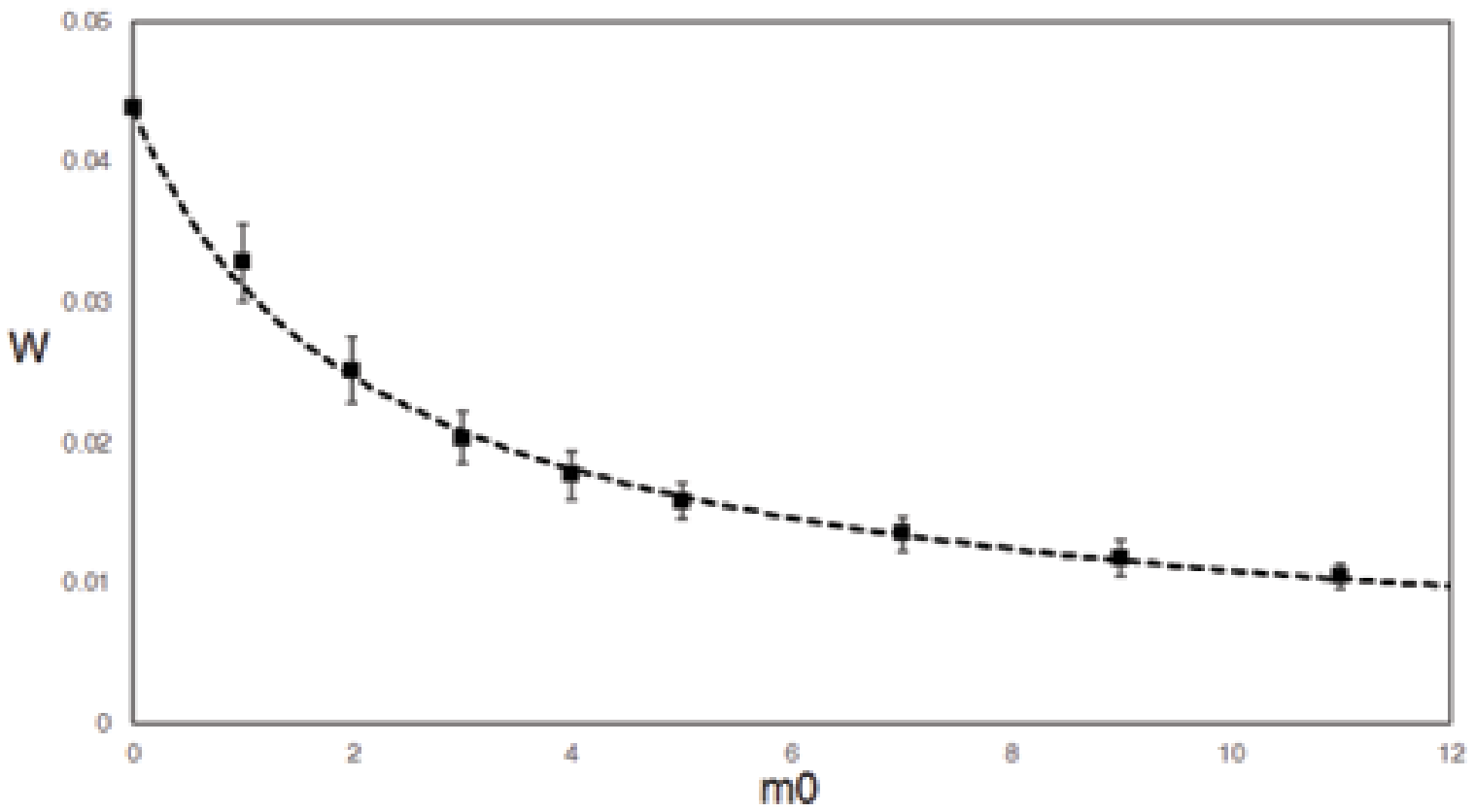}
\caption{Localization width $W$ as a function of $V_0$ for random potential (upper) and $m_0$ for random mass (lower). Nonlinear fits by Eq.\ref{fittingequation} (dashed lines) with parameters in Table \ref{fittingparam} are also shown on each figure.}
\label{localize}
\end{figure}

For the simulations, we calculate the time-evolution with the initial spinor (\ref{spreading_spinor})
\begin{equation}
\label{spreading_spinor}
\Psi(x)=\frac{\exp\left(-\frac{x^2}{4(0.04)^2}\right)}{\sqrt{0.04\sqrt{2\pi}}}
\left(
\begin{array}{c}
1\\0
\end{array}
\right)
\end{equation}
with $\langle m\rangle=500$ and $\langle V\rangle=0$.  We devided our simulations into two parts: random potential and random mass. For random potential, we increased the random strength $V_0$ with values 0, 1, 2, 4, 6, 8, 10, and 12, each with 100 samples. For random mass, the random strength $m_0$ are set to 0, 1, 2, 3, 4, 5, 7, 9, and 11, each with 100 samples. Each random value of $V$ or $m$ are spatially spaced with the length of numerical lattice constant $d_x$. The results of localization width $W[p(x)]$ (\ref{localization_width}) for each $V_0$ and $m_0$ as a function of time are shown in Fig.\ref{evol}. The values of $W$ at $t=0.71064$ are also shown in Fig.\ref{localize} and are also fitted (with weight $\sigma^{-2}[W]$) to the function
\begin{equation}
W(s)=(as+b)^{-\nu}
\label{fittingequation}
\end{equation}
with $s$ being either $V_0$ or $m_0$. The fitting results are shown in Table \ref{fittingparam}

\Table{\label{fittingparam}Fitting parameters and goodness of fit for Eq.\ref{fittingequation}} 
\br
&Random potential&Random Mass\\
\mr
$a$&30.08&56.9\\
$b$&52.42&89.07\\
$\nu$&0.7897&0.6965\\
$R^2$&0.9998&0.9998\\
\br
\endTable

\begin{figure}[H]
\centering
\includegraphics[scale=0.7]{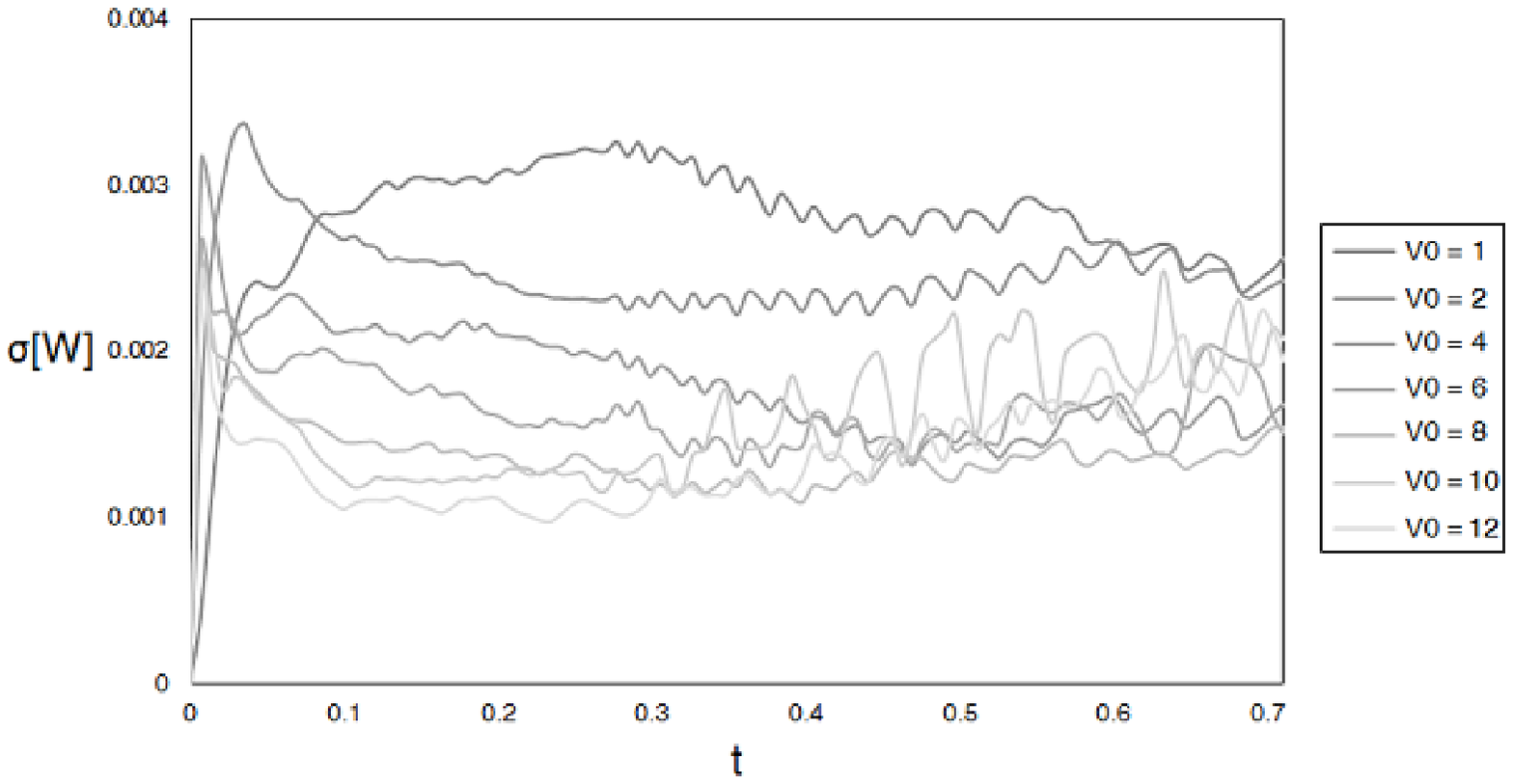}\\
\includegraphics[scale=0.7]{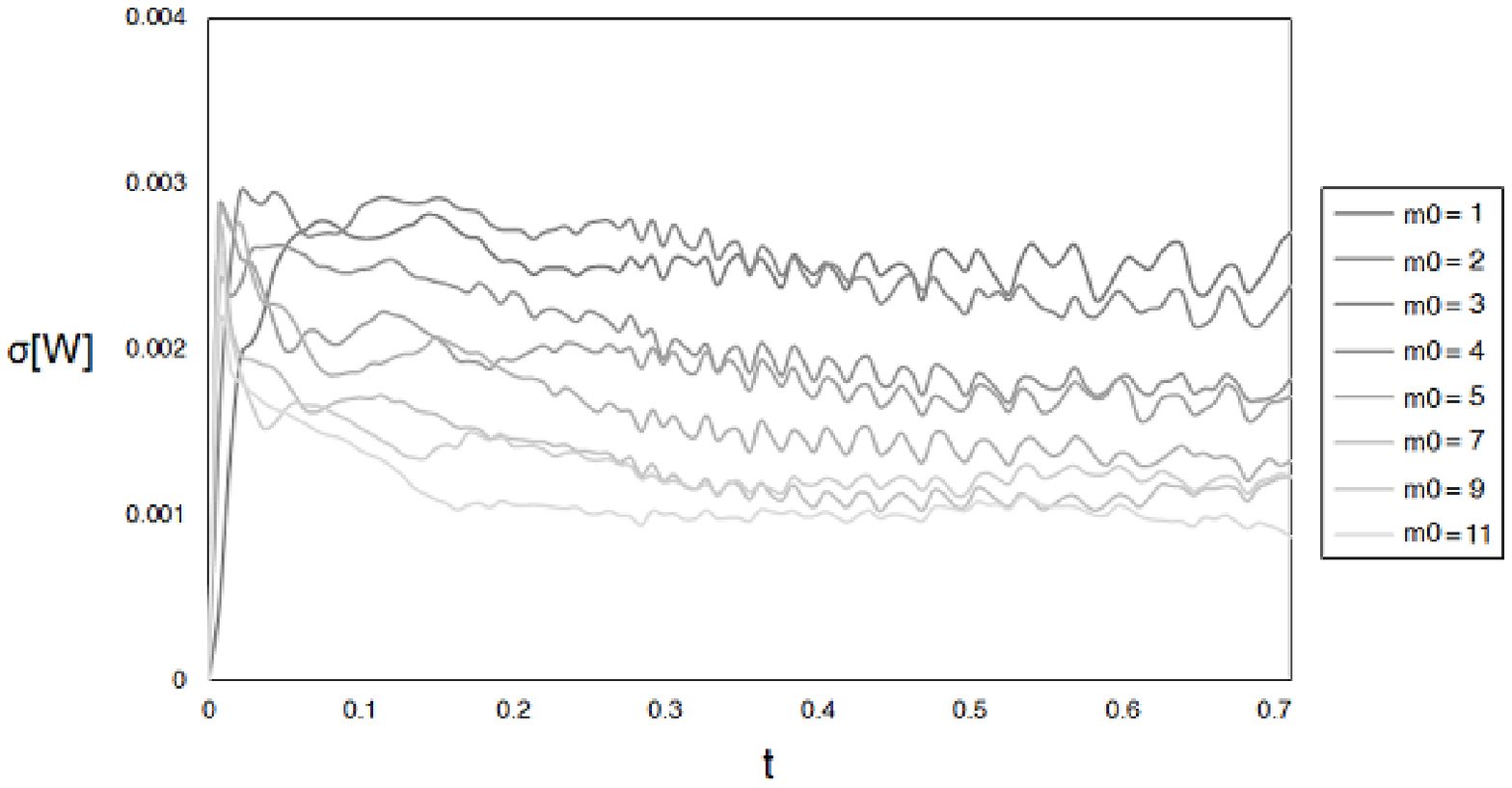}
\caption{The standard deviation of $W$ in Fig \ref{evol}}
\label{evolErr}
\end{figure}

Fig.\ref{evol} describe the time evolution of localization width $W$ as a function of time for each random strength. The dash lines represent a free particle case which disperse quadratically with time and can be approximated by Eq.\ref{width_dispersion}. After switching up the randomness, dispersion profiles clearly indicate localization. Fig.\ref{evolErr} show the absolute statistical errors of Fig.\ref{evol}. After some finite time, the widths converge to the specific values that depend on the random strength, as depicted in Fig.\ref{localize}. Notice that oscillations of $W$ and there errors emerge at $t>0.3$ with the same frequency. All of these are the oscillation of wave packets in the local finite well with the frequency depending on the well width $d_x$ \cite{PhysRevA.55.4526}.

\section{Discussion and Conclusion}
We have investigated the spread of the Gaussian wave packet from the time evolution of the Dirac equation. For the free Dirac equation, the results are obtained analytically and numerically. They are consistent with previous results by different methods. Interesting well-known behaviors such as the Zitterbewegung are evident. For the Dirac equation with random potential or mass, we employ the numerical algorithm to investigate the probability profiles of the displacement distribution when the potential is uniformly distributed. We observe that the width of the Gaussian wave function decreases with power law of order $o(s^{-\nu})$, with $\frac{1}{2} < \nu <1$, as the randomness strength $s$ of both potential and mass increases. While random mass and potential configurations lead to similar behavior, localization for random potential is slightly but significantly larger than localization with random mass. This suggests an onset of localization, but it is weaker than Anderson localization.

{\ack
A. Yosprakob would like to thank Sri-Trang Thong Scholarship, Faculty of Science, Mahidol University for financial support to study at Department of Physics, Faculty of Science, Mahidol University, and opportunities to conduct summer research overseas. S. Suwanna is grateful to the Faculty of Science, Mahidol University, for research funding for young researchers Grant No. A33/2554.
}

\printbibliography

\end{document}